\documentclass[a4paper, 10pt]{article}
\usepackage{graphicx}
\usepackage{url}

\renewcommand{\rmdefault}{cmss} 
\fontfamily{\rmdefault}
\selectfont
\date{}
\begin{document}

\begin{center}
{\bf{\huge{INPOP10a}}} 
\end{center}
\begin{center}
{\bf{\large{by A. Fienga$^{1,2}$,  H. Manche$^{1}$, P. Kuchynka$^{1}$, J. Laskar$^{1}$, M. Gastineau$^{1}$.}}}\\
\vspace{0.35cm}
September 21, 2010 \\
\vspace{0.35cm}
  {\small{$^{1}$ Astronomie et Syst\`emes Dynamiques, IMCCE-CNRS UMR8028, Paris, France}} \\
  {\small{$^{2}$ Observatoire de Besan\c con, UTINAM-CNRS UMR6213, Besan\c con, France}}
\end{center}

The INPOP10a version has several improvements in the fitting process, the data sets used in the fit and in the general features of the solution. No big change was brought in the dynamics.  As a consequence of these changes, the extrapolation capabilities of INPOP10a are improved compared to INPOP08, especially for the Earth, Mars, Mercury and Saturn orbits.
As for INPOP08, INPOP10a provides to the user, positions, velocities of the planets and the moon and TT-TDB chebychev polynomials at \url{http://www.imcce.fr/inpop}.

\section{The Data sets}
Several data sets have been added since INPOP08. The global distribution of the data used for the INPOP fit has changed its balance compared to INPOP06: now, more than 56$\%$ of the planetary data used in the fit are deduced from tracking data of spacecrafts including range deduced from tracking, VLBI angular positions and flyby normal points.
The statistics of the obtained postfit residuals are presented in table \ref{omctab} and \ref{Tab_residus_LLR_I10a}. Comments are given in section 4. Postfit residuals are plotted in figures \ref{obs1}, \ref{obs2}, \ref{obs3}, \ref{obs4}, and \ref{Fig_residus_LLR_I10a}, 

\subsection{Mercury}
\begin{itemize}
\item Two normal points deduced from the Mariner tracking data in 1974 and 1975 have been provided by JPL (Folkner 2010). These points have been included in the JPL DE solutions but not distributed. They bring constraints only in range with an accuracy of about 150 meters. This is an improvement of a factor 5 compared to the previous direct radar range observations of the Earth-Mercury distances.
\item Based on planetary ephemerides refitted by the Messenger navigation team to the tracking data of the Messenger mission during its flybys of Mercury (NAVPE004, NAVPE005 and NAVPE006 as provided by the SPICE NAIF) , we extracted 3 normal points giving 3 very accurate geocentric positions of Mercury. These positions are given in geocentric ($\alpha$,$\delta$,range) with an accuracy of about 1.5 mas, 2.5 mas and 5 meters respectively.
\end{itemize}
These 5 points change drastically our knowledge of the Mercury orbit. Until now, only direct radar ranging on the Mercury surface were avaliable with an accuracy of about 800 meters. 

\subsection{Mars and Venus}
Like with INPOP08,  tracking data of MEX and VEX misions are  provided by ESA (Morley 2009, Morley 2010) based on a procedure described in Fienga et al. (2009). For extrapolation tests, we decided to keep apart the last 6 months of MEX and VEX data in order to test the extrapolation capabilities of INPOP10a (see section \ref{extrap}).

\subsection{Saturn}
To the Cassini normal points provided by JPL over the 2005 to 2007 period and used in INPOP08, are added VLBI observations of the spacecraft published by (Jones et al. 2010). These new data are spread over a complementary period of time: 2 points before 2005 and 8 positions obtained from 2007 to 2009.5. They provide very accurate ICRF geocentric ($\alpha$,$\delta$) with accuracies better than 1 mas.

\subsection{Jupiter, Uranus and Neptune}
Flybys data of the 3 planets obtained during several missions (Pionneer 10 and 11, Viking 1 and 2, Ulysses and Cassini) are also added, provided by Folkner (2009). The mean accuracy for the Jupiter flybys in range is about 2 kilometers and the data are spread over 25 years. The range flybys for Uranus and Neptune are unique points with an accuracy of about several meters. 
These observations improve the estimations of the geocentric distances to the outer planets while no observation of that type were used in INPOP06 and INPOP08 adjustements.
New optical data obtained from 2000 to 2008 by the United States Naval Observatory Flagstaff Station using the Flagstaff Astrometric Scanning Transit Telescope are also added for Uranus, Neptune and Pluto.

\subsection{Pluto}
Stellar occultations are taken into account in INPOP10a by the use of measured offsets in topocentric ($\alpha$,$\delta$).
These offsets have been provided by Sicardy (2009). The accuracy of these observations is about 45 mas for $\alpha$ and 20 mas for $\delta$. Such accuracy is at least a factor 2 better than the modern CCD observations.

\subsection{Moon}
Lunar Laser Ranging observations from 1969 to 2010 are taken into account, provided by several sites: Cerga (Grasse, France), Mc~Donald, MLRS1, MLRS2 (Fort Davis, Texas, USA), Haleakala (Hawaii, USA) and Apollo (New Mexico, USA). Their accuracy reaches today a few centimeters. Since INPOP08, observations from Apollo and Cerga have been added. The weights of recent Cerga's observations have been split into 2 periods (before and after 1995), as for MLRS1 ones (before and after 1985).

\section{The fitting procedure}

\subsection{Fit of the mass of the sun for a fixed AU}

In INPOP10a, the value of the AU is fixed to the value given in the IERS2003 convention. The GM of the sun is fitted to the observations with the initial conditions of planets, the densities of the asteroids and the oblatness coefficient $J_{2}$ of the sun. Values of the fitted parameters are given in table \ref{paramfit}.

\subsection{Fit of a new selection of asteroids}

The selection of asteroids modeled in INPOP10a is based on Kuchynka et al. (2010). 
The perturbations of 24635 asteroids have been taken into account using an averaging of their perturbations 
by a ring with fixed physical characteristics  for most of them and individual perturbations 
for 161 of them (see Kuchynka et al. 2010). 
Based on a study of the correlations between the asteroids, we found 30 asteroids among the most perturbing objects highly correlated with each others.
In order to decrease the uncertainties on the mass estimations, we fixed 15 asteroid masses to values 
well determinated by other methods and found in Baer (2010).
Besides these fixed values, we fit the 146 GMs of asteroids using a BVLS algorithm (Lawson and Hanson, 1995)  with large constraints on the densities.

\section{General features}
\subsection{IAU2009 Current Best Estimates (CBE)}
We used the mass of planets provided by the IAU 2009 CBE lists (Luzum 2010).

\subsection{Fit of parameters related to LLR observations}
More than 180 parameters have a significant effect on LLR computations, including initial conditions, potential coefficients, reflectors and stations coordinates, biases,... But because of large correlations or better estimations obtained by other techniques, only 59 of them have been fitted. They have been choosen so that the ratio between the formal error and the fitted value is less than 5\%. Values of parameters and formal errors are given in tables \ref{Tab_valeurs_parametres_LLR_dyn_I10a}, \ref{Tab_valeurs_parametres_LLR_I10a_reflecteurs} and \ref{Tab_valeurs_parametres_LLR_I10a_stations}.
Initial values for Moon's coefficients of potential are from LP150Q gravity model described in Konopliv (2001), and some of them are then fitted.
Initial values for Earth's coefficients of potential are from EGM96 (\url{http://cddis.nasa.gov/926/egm96/egm96.html}); $J_2$ and $J_3$ are then fitted.
Furthermore, two biases have been fitted:
\begin{itemize}
 \item the first one on Cerga's observations between December 1996 and June 1998. It was first noticed by Chapront-Touz\'e et al. (2000). Its fitted value is -0.6~ns.
 \item the second one is applied on Mc~Donald's observations, between October 1972 and August 1975. Its fitted value is 2.2~ns.
\end{itemize}

The final version of INPOP10a was obtained after an iterative process between planetary fit and Moon fit.

\section{Results and Applications}
\label{results}
The table \ref{omctab} gives the planetary postfit residuals obtained with INPOP08 and INPOP10a.
Some of the data sets used in the INPOP10a were not used in the INPOP08 adjustement: this explains some differences between the two columns of table \ref{omctab}.
Postfit residuals are plotted in figures \ref{obs1}, \ref{obs2}, \ref{obs3}, and \ref{obs4}. 
The figures \ref{Fig_residus_LLR_I10a} show LLR residuals for each station and table \ref{Tab_residus_LLR_I10a} gives their standard deviations.

\subsection{Asteroid masses}

In table 2 and figures 1 and 2 are presented the asteroid masses obtained with INPOP10a compared to values found in the literature. 
The major sources of comparisons are the values obtained with DE421 (Folkner et al. 2009), DE423 (Konopliv et al. 2010) and 
Baer (2010) with realistic errorbars estimated by Kuchynka (2010). 
On plots 1 and 2, the comparisons of the asteroid GMs are ranked by their impact on the Earth-Mars distances over the 1990 to 2010 period. 
It then appears clearly that the estimations for the most perturbing objects are quite consistent when the estimations of the weak perturbing objects show bigger discrepancies. 
This result is consistent with the planetary ephemerides method for asteroid mass computation. 
For the asteroids inducing perturbations up to 10 meters, the differences in GMs are usually below 1-sigma or very close to 1-sigma except for 52 Europa.
For this asteroid, which induced a maximum of 10 meters on the Earth-Mars distances, the errorbars are very large for all the determinations based on planetary ephemerides: we thus conclude to a bad determination of this mass based on the present interval of available data. 
In such case, one should have fixed this mass to the value estimated by other methods and 
especially by close-encounters as the one obtained by Baer et al (2008). 
This should be done for the next INPOP version. However, the good quality of the INPOP10a postfit and extrapolated residuals already tells us that such modifications would not change significantly the global behaviour of the ephemerides.

\subsection{Extrapolation tests}
\label{extrap}

Table \ref{omcextrap} gives the residuals obtained by extrapolation of the two ephemerides INPOP08 and INPOP10a out of their fit interval. These residuals are plotted in figure 8. 
INPOP10a is fitted over the [1914.2:2009.7] time interval when INPOP08 was fitted over the interval [1914.2:2008.2].
For Venus, the extrapolation residuals are the same for INPOP08 and INPOP10a.
For Mars, we first compare the extrapolation residuals of INPOP10a and of INPOP08 over the same time interval, from 2009.7 to 2010.35. Over this period, the residuals of INPOP10a are 9 times smaller than those obtained with INPOP08 (see table \ref{omcextrap}). This is obviously induced by the prolongation of the fitting interval and the addition of new data sets.
In order to really test the improvement of the INPOP extrapolation capabilities, we compare the residuals obtained 6 months after the end of the fit period, on [2008.2:2008.8] for INPOP08 and over [2009.8:2010.2] for INPOP10a. Over these two intervals, the INPOP10a residuals have a clear linear drift of 20 meters/yr while the INPOP08 residuals show a more quadratic trend, as one can see in figure 9. This pure linear trend of INPOP10a extrapolation residuals can easily be removed (top right-hand plot of figure 9). This is an improvement compared to the residuals obtained with INPOP08. Indeed even after the correction of the linear trend, a quadratic signal is still remaining in the INPOP08 extrapolation residuals. 
This new behaviour of INPOP10a extrapolation capabilities could be induced by the changes in the fit of the asteroid masses.

\subsection{Determination of sun $J_{2}$ and PPN $\beta$}
 
Table \ref{paramfit} gives the obtained values for the mass of the Sun, the sun $J_{2}$ and the interval of
sensitivity of data to modification of PPN $\beta$. Based on Fienga et al. (2010), we realized several fits for different values of $\beta$ with a simultaneous fitting of initial conditions of planets, mass of the sun and asteroid densities.
The given interval corresponds to values of $\beta$ inducing changes in the postfit residuals below 5$\%$. As one can see in table \ref{paramfit}, we obtain with INPOP10a an improved estimation of $\beta$ mainly due to the use in the fit of the Mariner and Messenger normal points of Mercury.


\section{Work in progress}

A complementary work is in progress to improve the physical interpretation of the fitted masses of asteroids.
New estimations of supplementary advances of perihelia of planets (especially Saturn) will be computed rapidely as well as combined estimations of the PPN parameters $\beta$ and $\gamma$.

\section{References}

Baer, J., 2010, \url{http://home.earthlink.net/~jimbaer1/astmass.txt} \\

\noindent Baer, J.; Milani, A.; Chesley, S.; Matson, R. D, 2008, {\it{IAn Observational Error Model, and Application to Asteroid Mass Determination}}, American Astronomical Society, DPS meeting 40, 52.09; Bulletin of the American Astronomical Society, Vol. 40, p.493.\\

\noindent Chapront-Touz\'e, M.; Chapront, J.; Francou, G., 2000, {\it{Determination of UT0 with LLR observations}}, Journ\'ees 1999 - syst{\`e}mes de r{\'e}f{\'e}rence spatio-temporels, p. 217-220.\\

\noindent Fienga, A., .; Laskar, J.; Morley, T.; Manche, H.; Kuchynka, P.; Le Poncin-Lafitte, C.; Budnik, F.; Gastineau, M.; Somenzi, L., 2009, {\it{INPOP08, a 4-D planetary ephemeris: from asteroid and time-scale computations to ESA Mars Express and Venus Express contributions}}, Astronomy and Astrophysics, 507, 1675.\\

\noindent Fienga, A.; Laskar, J.; Kuchynka, P.; Le Poncin-Lafitte, C.; Manche, H.; Gastineau, M., 2010, {\it{Gravity tests with INPOP planetary ephemerides}}, Proceedings of the IAU Symposium "Relativity in Fundamental Astronomy: Dynamics, Reference Frames, and Data Analysis", Volume 261, p. 159-169.\\

\noindent Folkner, W.M., 2010, private communication.\\

\noindent Folkner, W.M., 2009, private communication.\\

\noindent Jones, Dayton L.; Fomalont, E.; Dhawan, V.; Romney, J.; Lanyi, G.; Border, J.; Folkner, W.; Jacobson, R., 2010, {\it{Astrometric Observations of Cassini with the VLBA: The First Eight Epochs}}, American Astronomical Society, AAS Meeting 215, 448.13; Bulletin of the American Astronomical Society, Vol. 42, p.456.\\

\noindent Konopliv et al., 2010, in press or submitted ?\\

\noindent Konopliv, A.~S.; Asmar, S.~W.; Carranza, E.; Sjogren, W.~L.; Yuan, D.~N., 2001, {\it{Recent Gravity Models as a Result of the Lunar Prospector Mission}}, Icarus, 150, p. 1-18\\

\noindent Kuchynka, P., Laskar, J., Fienga, A., Manche, H., 2010., {\it{A ring as a model of main-belt in planetary ephemerides.}}, Astronomy and Astrophysics, 514, A96.\\
\noindent Kuchynka, P., 2010, Paris Observatory PhD,  {\it{}}\\

\noindent Lawson, C., Hanson, R., 1995, in {\it{Solving Least Squares Problems, Revised edition,}}, eds. SIAM\\

\noindent Luzum, B.J., 2010, {\it{System of Astronomical Constants}}, Proceedings of the IAU Symposium "Relativity in Fundamental Astronomy: Dynamics, Reference Frames, and Data Analysis", Volume 261, p. 878.\\

\noindent Morley, T., 2010, private communication.\\

\noindent Morley, T., 2009, private communication.\\

\noindent Sicardy, B., 2009, private communication.\\

\begin{table}
\caption{Statistics of the residuals obtained after the INPOP10a fit. For comparisons, means and standard deviations of residuals obtained with INPOP08 are given. }
\begin{center}
\begin{tabular}{l l l | c | c }
\hline
\small{Planet} & & & \small{INPOP08} & \small{INPOP10a}\\
\small{Type of Data} & Nbr& Time & & \\
 & & Interval & mean $\pm$ 1$\sigma$& mean $\pm$ 1$\sigma$\\
\hline
\small{Mercury} & & & & \\
\small{Direct range [m]} & 462 & 1965-2000 & {30 $\pm$ 842} &  {7 $\pm$ 866}\\
\small{Mariner range [m]}& 2 & 1974-1975& {-1000 $\pm$ 305} & {-28 $\pm$ 85}\\
\small{Messenger Mercury flybys} & & & & \\
\small{Messenger ra [mas]} & 3 & 2008-2009 & { 1.1 $\pm$ 0.7 } & {0.4 $\pm$ 1.2}\\
\small{Messenger de [mas]} & 3 & 2008-2009 & {2.0 $\pm$ 1.9} & {1.9 $\pm$ 2.1}\\
\small{Messenger range [m]} & 3 & 2008-2009 & {52 $\pm$ 619} & {-0.6 $\pm$ 1.9}\\
\hline
\small{Venus } & & & & \\
\small{ Direct range [km]} & 489 & 1965-2000 & { 0.5 $\pm$ 2.3} &  {0.5 $\pm$ 2.2}\\
\small{ VEX range [m]} & 22145 & 2006-2010 & {1.6 $\pm$ 4.4} &  {-0.2 $\pm$ 3.9}\\
\small{ VLBI [mas]} & 22 & 1990-2007 & {2 $\pm$ 2} &  { 2 $\pm$ 2.5 }\\
\hline
\small{Mars range [m] } & & & & \\
\small{MGS} & 10474 & 1998-2008 & \small{-0.9 $\pm$ 1.6} &     \small{0.5 $\pm$ 1.9} \\
\small{MEX} & 24262 & 2006-2010 & {-3.5 $\pm$ 2.0} &    {0.0 $\pm$ 1.7} \\
\small{Path} & 90 & 1997 & {6.8 $\pm$ 12.5} &    {-5.0 $\pm$ 5.0} \\
\small{Vkg} & 1256 & 1976-1982 & {-27.4 $\pm$ 19.0} &    {-5.7 $\pm$ 35.0} \\
\small{Mars VLBI [mas]} & 96 & 1989-2007 & {0.5 $\pm$ 0.5} &  { -0.0 $\pm$ 0.4}\\
\hline
\small{Jupiter flybys} & & & & \\
ra [mas] & 5 & 1974-2000 &{48.0 $\pm$ 40.0}  &  {6 $\pm$ 5} \\
de [mas] & 5 &  1974-2000 &{-10.0 $\pm$ 50} &  {-13 $\pm$ 18} \\
\small{range [km]} & 5 & 1974-2000 &  {-27$\pm$ 55} &  {-0.6 $\pm$ 1.6} \\
\small{Jupiter VLBI} [mas] & 24 & 1996-1997&  {4 $\pm$ 11} & {0.2 $\pm$ 11}\\
\small{Jupiter Optical} & & \\
ra [mas] & 6216 & 1914-2008 &  {20 $\pm$ 304} & {-26 $\pm$ 304}\\
de [mas] &  6082 & 1914-2008 &  {-44 $\pm$ 313} & {-54 $\pm$ 303}\\
\hline
\small{Saturn Cassini} & & & & \\
ra [mas] & 31 & 2004-2007 & {1.5 $\pm$ 4}  &  {0.7 $\pm$ 4} \\
de [mas] &  31 & 2004-2007 & {7.0 $\pm$ 7} &  {6.5 $\pm$ 7} \\
\small{range [m]} & 31 & 2004-2007 &  {0.5 $\pm$ 22} &  {0.0 $\pm$ 17} \\
\small{Saturn VLBI Cassini} & & & & \\
ra [mas] & 10 & 2004-2009 & {0.3 $\pm$ 0.7}  &  {0.0 $\pm$ 0.6} \\
de [mas] & 10 & 2004-2009 &  {-1.2 $\pm$ 2.0} &  {0.1 $\pm$ 0.4} \\
\small{Saturn Optical} & & & & \\
ra [mas] & 7824 & 1914-2008 &  {-16 $\pm$ 305} & {-16 $\pm$ 305}  \\
de [mas] & 7799 & 1914-2008 &  {-7 $\pm$ 276} & {-9 $\pm$ 276}   \\
\hline
\small{Uranus flybys} &  & & & \\
ra [mas] & 1 & 1986 &{-90}  &  {-30} \\
de [mas] & 1 &  1986 &{-36} &  {-7} \\
\small{range [km]} & 1 &  &  {1139} &  {0.080} \\
\small{Uranus Optical} & & & & \\
ra [mas] & 4145 & 1914-2008 &  { -44 $\pm$ 278 } & { -27 $\pm$ 290}  \\
de [mas] & 4130  & 1914-2008 &  { -38 $\pm$ 339 } & { -11 $\pm$ 338}   \\
\hline
\small{Neptune flybys} & & & & \\
ra [mas] & 1 & 1989 &{-88}  &  {-11} \\
de [mas] & 1 &  1989 &{-48} &  {-10} \\
\small{range [km]} & 1 &  &  {2305} &  {0.004} \\
\small{Neptune Optical} & & & & \\
ra [mas] & 4340 & 1914-2008 &  { -32 $\pm$ 282} & { 2 $\pm$ 281}  \\
de [mas] & 4320 & 1914-2008 &  { -36 $\pm$ 335} & { 2 $\pm$ 330}   \\
\hline
\small{Pluto occultation} & & &&\\
ra [mas] & 13 & 2005-2009 &  {-6 $\pm$ 46} & {-1 $\pm$ 47}  \\
de [mas] & 13 & 2005-2009 &  {16 $\pm$ 30} & {-2 $\pm$ 19}   \\
\small{Pluto Optical} & & &&\\
ra [mas] & 2449 & 1914-2008 &  {353 $\pm$ 926} & {38 $\pm$ 629}  \\
de [mas] & 2463 & 1914-2008 &  {-22 $\pm$ 524} & {17 $\pm$ 536}   \\
\hline
\end{tabular}
\label{omctab}
\end{center}
\end{table}

\begin{table}
\caption{Asteroids masses found in the recent literature and compared to the values estimated in INPOP10a. (F) is for fixed value.}
\begin{center}
\begin{tabular}{l l l l l}
\hline
\small{Asteroids} & \small{INPOP10a} & \small{Baer 2010} & \small{DE421} & \small{DE423}\\
IAU& $10^{12} \times M_{\odot}$ & $10^{12} \times M_{\odot}$ & $10^{12} \times M_{\odot}$ & $10^{12} \times M_{\odot}$ \\
\hline
1 & 475.836 $ \pm $ 2.849 & 475.500 $ \pm $ 4.755 & 468.488 $ \pm $ 46.849 & 467.900 $ \pm $ 3.250 \\
2 & 111.394 $ \pm $ 2.808 & 106.000 $ \pm $ 1.060 & 100.979 $ \pm $ 10.098 & 103.440 $ \pm $ 2.550 \\
4 & 133.137 $ \pm $ 1.683 & 133.070 $ \pm $ 0.266 & 132.835 $ \pm $ 13.284 & 130.970 $ \pm $ 2.060 \\
7 & 7.772 $ \pm $ 1.142 & 6.860 $ \pm $ 1.029 & 5.998 $ \pm $ 0.600 & 5.530 $ \pm $ 1.320 \\
324 & 4.692 $ \pm $ 0.379 &  & 4.980 $ \pm $ 0.498 & 5.340 $ \pm $ 0.990 \\
3 & 11.604 $ \pm $ 1.313 & 13.400 $ \pm $ 2.278 & 11.573 $ \pm $ 1.157 & 12.100 $ \pm $ 0.910 \\
6 & 7.084 $ \pm $ 1.212 & 6.460 $ \pm $ 0.969 & 4.558 $ \pm $ 0.456 & 6.730 $ \pm $ 1.640 \\
8 & 4.072 $ \pm $ 0.631 & 4.260 $ \pm $ 0.980 & 1.778 $ \pm $ 0.178 & 2.010 $ \pm $ 0.420 \\
9 & 5.700 (F) & 5.700 $ \pm $ 1.425 & 4.272 $ \pm $ 0.427 & 3.280 $ \pm $ 1.080 \\
10 & 44.500 (F) & 44.500 $ \pm $ 0.890 & 40.416 $ \pm $ 4.042 & 44.970 $ \pm $ 7.760 \\ 
11 & 1.886 $ \pm $ 1.029 & 3.090 $ \pm $ 0.989 & 2.682 $ \pm $ 0.268 &  \\
13 & 8.200 (F) & 8.200 $ \pm $ 1.640 & 3.104 $ \pm $ 0.310 &  \\
14 & 4.130 (F) & 4.130 $ \pm $ 0.991 & 2.622 $ \pm $ 0.262 & 1.910 $ \pm $ 0.810 \\
15 & 18.856 $ \pm $ 1.617 & 15.700 $ \pm $ 0.942 & 12.342 $ \pm $ 1.234 & 14.180 $ \pm $ 1.490 \\
16 & 11.212 $ \pm $ 5.174 & 11.000 $ \pm $ 0.990 & 16.825 $ \pm $ 1.682 & 12.410 $ \pm $ 3.440 \\
18 & 1.845 (F) &  & 2.012 $ \pm $ 0.201 &  \\
19 & 6.380 (F) & 6.380 $ \pm $ 1.021 & 3.489 $ \pm $ 0.349 & 3.200 $ \pm $ 0.530 \\
20 & 2.850 (F) & 2.850 $ \pm $ 0.998 & 2.193 $ \pm $ 0.219 &  \\
21 & 1.275 $ \pm $ 1.170 &  & 1.047 $ \pm $ 0.105 &  \\
24 & 2.826 $ \pm $ 1.902 & 5.670 $ \pm $ 2.155 & 3.036 $ \pm $ 0.304 &  \\
25 & 0.002 $ \pm $ 0.002 &  & 0.301 $ \pm $ 0.030 &  \\
28 & 4.652 $ \pm $ 0.990 &  & 1.243 $ \pm $ 0.124 &  \\
29 & 5.920 (F) & 5.920 $ \pm $ 1.006 & 6.826 $ \pm $ 0.683 & 7.420 $ \pm $ 1.490 \\
31 & 3.130 (F) & 3.130 $ \pm $ 1.002 & 8.582 $ \pm $ 0.858 &  \\
41 & 9.213 $ \pm $ 2.631 &  & 3.971 $ \pm $ 0.397 & 4.240 $ \pm $ 1.770 \\
42 & 1.854 $ \pm $ 1.070 &  & 0.693 $ \pm $ 0.069 &  \\
45 & 2.860 (F) & 2.860 $ \pm $ 0.057 & 2.991 $ \pm $ 0.299 &  \\
51 & 0.726 $ \pm $ 0.421 &  & 1.085 $ \pm $ 0.108 &  \\
52 & 42.304 $ \pm $ 8.056 & 8.290 $ \pm $ 0.995 & 10.202 $ \pm $ 1.020 & 11.170 $ \pm $ 8.400 \\
60 & 0.402 $ \pm $ 0.375 &  & 0.158 $ \pm $ 0.016 &  \\
63 & 2.022 $ \pm $ 1.685 &  & 0.769 $ \pm $ 0.077 &  \\
65 & 7.173 $ \pm $ 4.256 & 8.930 $ \pm $ 0.982 & 5.229 $ \pm $ 0.523 &  \\
78 & 3.232 $ \pm $ 2.319 &  & 0.640 $ \pm $ 0.064 &  \\
94 & 15.847 $ \pm $ 11.522 &  & 3.119 $ \pm $ 0.312 &  \\
105 & 1.109 (F) &  & 0.663 $ \pm $ 0.066 &  \\
107 & 18.205 $ \pm $ 4.647 & 5.630 $ \pm $ 0.169 &  &  \\
130 & 11.152 $ \pm $ 8.032 & 3.320 $ \pm $ 0.199 &  &  \\
135 & 0.917 $ \pm $ 0.884 &  & 0.588 $ \pm $ 0.059 &  \\
139 & 5.896 $ \pm $ 3.315 &  & 1.417 $ \pm $ 0.142 &  \\
145 & 2.266 (F) &  & 1.138 $ \pm $ 0.114 &  \\
187 & 2.484 $ \pm $ 1.075 &  & 0.791 $ \pm $ 0.079 &  \\
192 & 0.719 (F) &  & 0.806 $ \pm $ 0.081 &  \\
194 & 8.803 $ \pm $ 2.921 &  & 1.371 $ \pm $ 0.137 &  \\
216 & 0.564 $ \pm $ 0.460 &  & 2.253 $ \pm $ 0.225 &  \\
253 & 0.904 $ \pm $ 0.650 & 0.052 $ \pm $ 0.002 &  &  \\
337 & 0.543 $ \pm $ 0.080 &  & 0.249 $ \pm $ 0.025 &  \\
344 & 0.342 $ \pm $ 0.188 &  & 0.859 $ \pm $ 0.086 &  \\
354 & 2.451 (F) &  & 2.464 $ \pm $ 0.246 &  \\
372 & 4.443 (F) &  & 2.675 $ \pm $ 0.267 &  \\
419 & 0.997 $ \pm $ 0.554 &  & 0.769 $ \pm $ 0.077 &  \\
451 & 20.978 $ \pm $ 14.797 &  & 4.596 $ \pm $ 0.460 &  \\
488 & 6.234 $ \pm $ 5.539 &  & 1.236 $ \pm $ 0.124 &  \\
511 & 19.903 $ \pm $ 4.068 & 22.000 $ \pm $ 1.100 & 12.342 $ \pm $ 1.234 & 8.580 $ \pm $ 5.930 \\
532 & 2.895 $ \pm $ 0.759 &  & 6.676 $ \pm $ 0.668 & 4.970 $ \pm $ 2.810 \\
554 & 1.575 $ \pm $ 1.277 &  & 0.332 $ \pm $ 0.033 &  \\
704 & 18.600 (F) & 18.600 $ \pm $ 1.116 & 18.565 $ \pm $ 1.857 & 19.970 $ \pm $ 6.570 \\
747 & 6.032 $ \pm $ 2.306 &  & 1.477 $ \pm $ 0.148 &  \\
804 & 2.513 $ \pm $ 1.820 & 2.020 $ \pm $ 1.010 &  &  \\
\hline
\end{tabular}

\label{comparmass}
\end{center}
\end{table}

\begin{figure}
\begin{center}
\includegraphics[scale=0.3]{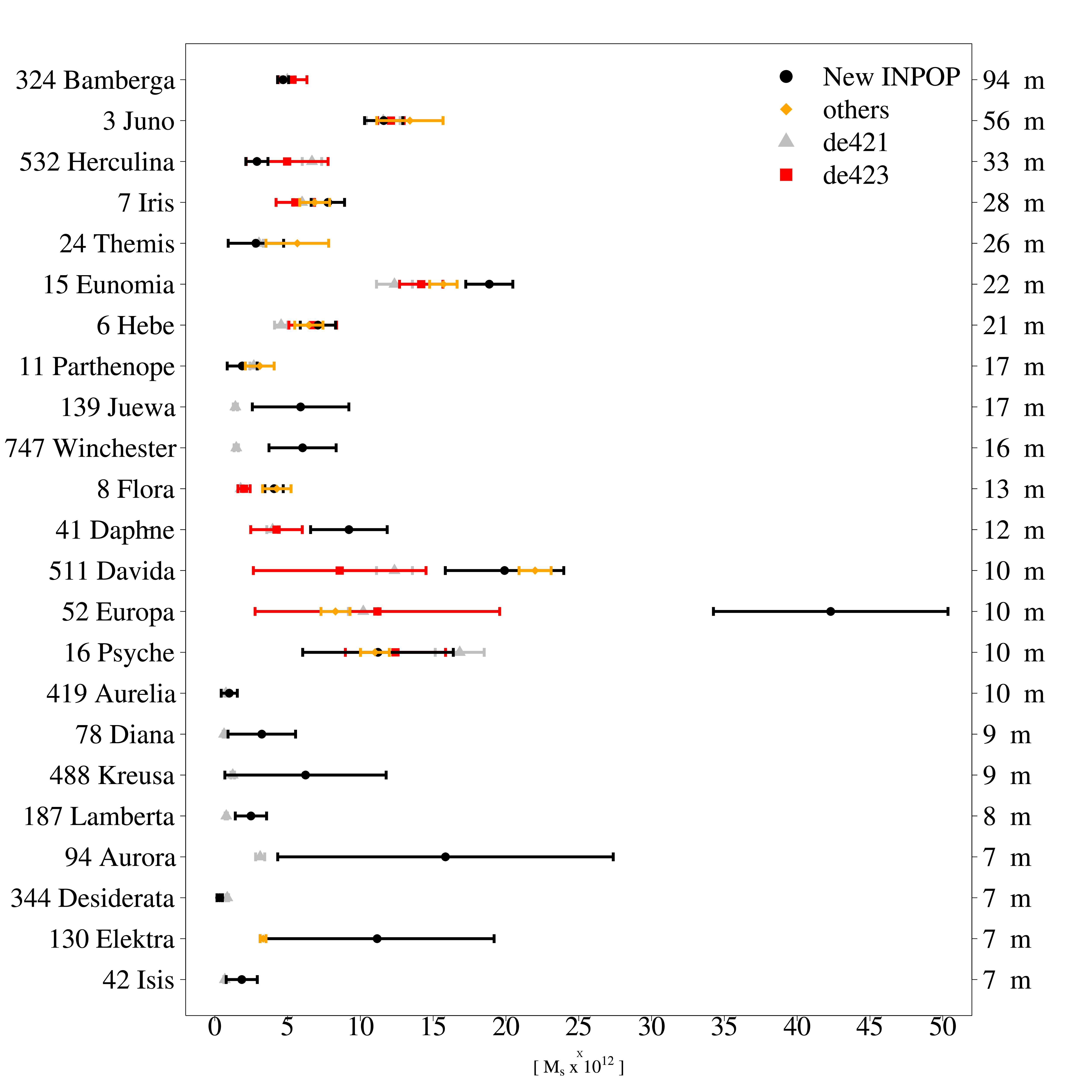}
\caption{Comparisons of the first 23 asteroid masses given in $10^{12}$ solar mass estimated by different authors and 
ranked by their impact on the Earth-Mars distances over 1990 to 2010. The mass estimations of the 3 biggest asteroids and perturbers Ceres, Pallas and Vesta are given in table 2. "others" stands for Baer (2010).}
\label{obs1}
\end{center}
\end{figure}

\begin{figure}
\begin{center}
\includegraphics[scale=0.3]{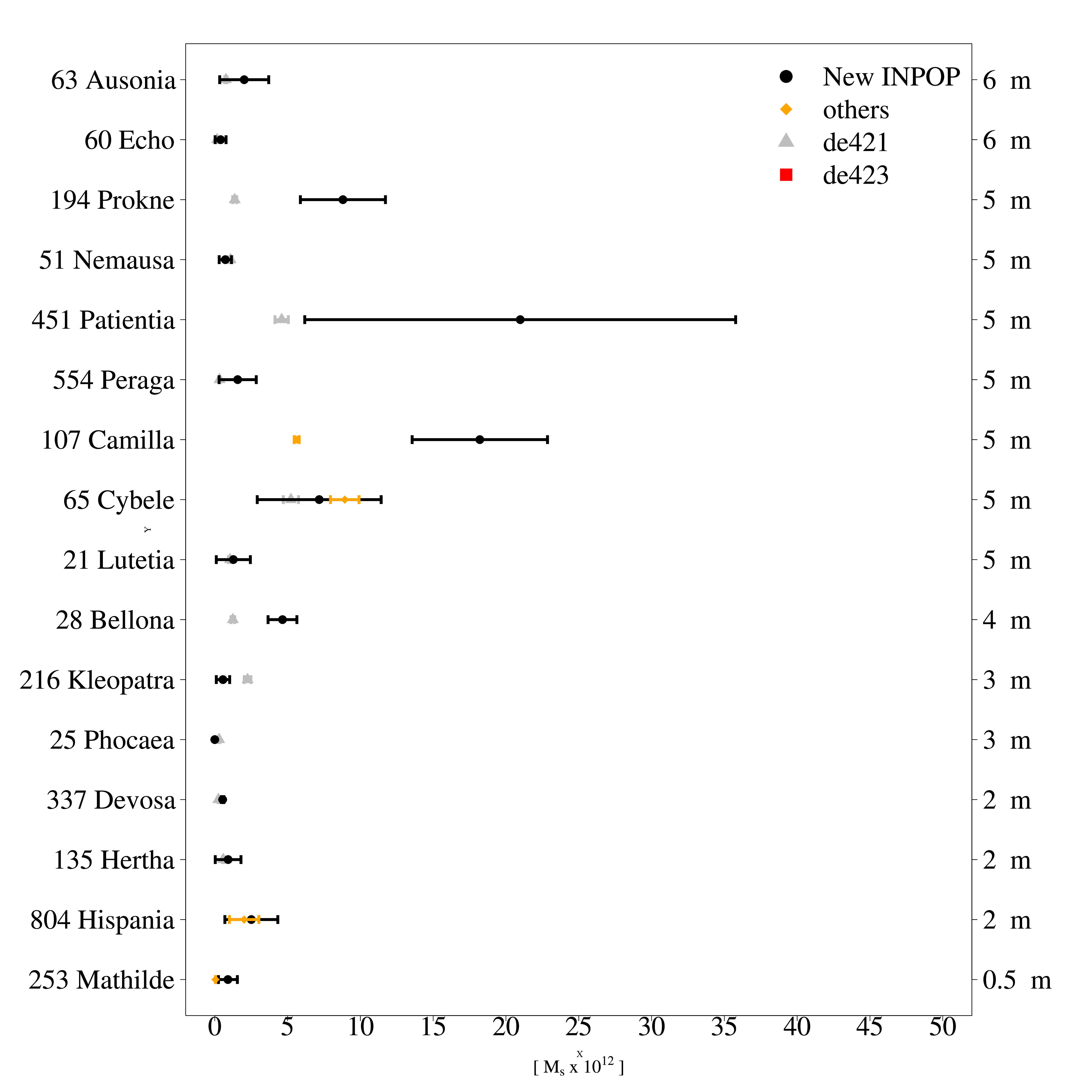}
\caption{Comparisons of the other asteroid masses given in $10^{12}$ solar mass estimated by different authors and ranked by their impact on the Earth-Mars distances over 1990 to 2010. "others" stands for Baer (2010).}
\label{obs2s}
\end{center}
\end{figure}

\begin{table}
\caption{Values of paramaters obtained by the fit of INPOP08 and INPOP10a to observations. In INPOP08, the mass of the sun was not fitted directly but deduced from the fitted value of AU (Fienga et al. 2009). The (F) indicates that the marked values are fixed in the fit. The published uncertainty  (Standish 1998) of the INPOP08 GM${\odot}$ fixed value is $\pm$ 50 km$^{3}.$ s$^{-2}$. The equivalent value of AU deduced from the estimation of the GM${\odot}$ in INPOP10a is given in line 6. The INPOP $\sigma$ for the Earth-Moon mass ratio EMRAT, the J2$_{\odot}$ and the PPN parameter $\beta$ are the limit of sensitivity of the observations to the tested parameter. For the AU and the GM${\odot}$, the $\sigma$ are deduced from the least squares fit.}
\begin{center}
\begin{tabular}{r c c c}
\hline
& INPOP08 & INPOP10a & DE423\\
&  $\pm$ 1$\sigma$&  $\pm$ 1$\sigma$ & \\
\hline
EMRAT & (81.30054 $\pm$ 0.00005) & (81.3005700 $\pm$ 0.0000010) & (81.3005694 $\pm$ 0.0000015) \\
\\
J2$_{\odot}$ & (1.82 $\pm$ 0.47) $\times$ 10$^{-7}$ & (2.40 $\pm$ 0.25) $\times$ 10$^{-7}$ &\\
\\
($\beta$ -1) $\times$ 10$^{-4}$ & (0.75 $\pm$ 1.25)  & (0.25 $\pm$ 0.75) & (0.4 $\pm$ 2.4) \\
\\
\hline
\\
GM${\odot}$ [km$^{3}.$ s$^{-2}$]&  132712440017.98700 (F) & 132712440055 $\pm$ 1 & 132712440042 $\pm$ 10 \\
 & & \\
AU [m] & 149597870699.2 $\pm$ 0.11 & 149597870691.0 (F) & \\ 
 & & &\\
AU [m] from GM${\odot}$ & & 149597870704.9 $\pm$ 0.3 & 149597870700.0 $\pm$ 3 \\
\hline
\end{tabular}
\footnotetext[1]{fixed}
\label{paramfit}
\end{center}
\vspace{2cm}
\caption{Residuals obtained by extrapolation of INPOP08 and INPOP10a out from the fit interval.}
\begin{center}
\begin{tabular}{l l | c | c }
\hline
\small{Planet} & & \small{INPOP08} & \small{INPOP10a}\\
\small{Type of Data} & & & \\
 & & mean $\pm$ 1$\sigma$& mean $\pm$ 1$\sigma$\\
\hline
\small{Mars range [m]} & & & \\
\small{MEX} & 2009.7:2010.35 & {-18 $\pm$ 63} &    {11.3 $\pm$ 8.6} \\
\hline
\small{VEX} & 2009.7:2010.35 & { 17 $\pm$ 50} &  {18 $\pm$ 50}\\
\hline
\end{tabular}
\label{omcextrap}
\end{center}
\end{table}

\begin{table}
\caption{Masses of asteroids fitted during the fit of INPOP10a. The $\sigma$ is deduced from the least squares fit.}
\begin{center}
\begin{tabular}{l c c | l c c | l c c}
\hline
Asteroids & M & $\sigma$ & Asteroids & M & $\sigma$ & Asteroids & M & $\sigma$\\
IAU& $10^{12} \times M_{\odot}$ & $10^{12} \times M_{\odot}$ & & $10^{12} \times M_{\odot}$ & $10^{12} \times M_{\odot}$ & & $10^{12} \times M_{\odot}$ & $10^{12} \times M_{\odot}$\\
\hline
1 & 475.836 & 2.849 &	84 & 0.275 & 0.204 &	381 & 4.615 & 0.400 \\
2 & 111.394 & 2.808 &	89 & 4.181 & 2.083 &   386 & 11.825 & 0.210 \\
4 & 133.137 & 1.683 &	94 & 15.847 & 11.522 &	387 & 2.674 & 0.310 \\
7 & 7.772 & 1.142 &	105 & 1.109 & (F)&      404 & 1.718 & 1.522 \\
324 & 4.692 & 0.379 &	107 & 18.205 & 4.647 &	410 & 3.476 & 2.380 \\
3 & 11.604 & 1.313 &	112 & 0.990 & 3.410 &	416 & 1.643 & 1.560 \\
6 & 7.084 & 1.212 &	117 & 8.659 & 0.100 &	419 & 0.997 & 0.554 \\
8 & 4.072 & 0.631 &	126 & 0.237 & 2.910 &	420 & 7.417 & 0.460 \\
9 & 5.700 & (F) &   127 & 4.402 & 0.440 &	442 & 0.098 & 0.010 \\
10 & 44.500 & (F) & 128 & 4.583 & 2.772 &	444 & 7.992 & 5.581 \\
11 & 1.886 & 1.029 &	129 & 4.336 & 1.846 &	445 & 1.744 & 0.390 \\
12 & 3.774 & 1.679 &	130 & 11.152 & 8.032 &	449 & 0.791 & 0.703 \\
13 & 8.200 & (F) & 132 & 0.207 & 1.360 &      451 & 20.978 & 14.797 \\
14 & 4.130 & (F) & 135 & 0.917 & 0.884 &	455 & 0.596 & 0.060 \\
15 & 18.856 & 1.617 &	138 & 0.248 & 0.130 &	469 & 3.647 & 3.140 \\
16 & 11.212 & 5.174 &	139 & 5.896 & 3.315 &	471 & 6.359 & 0.360 \\
18 & 1.845 & (F) &	141 & 4.146 & 2.902 &	481 & 2.908 & 0.730 \\
19 & 6.380 & (F) &	144 & 4.566 & 2.977 &	485 & 0.686 & 0.220 \\
20 & 2.850 & (F) &	145 & 2.266 & (F) &	488 & 6.234 & 5.539 \\
21 & 1.275 & 1.170 &	147 & 6.185 & 0.240 &	491 & 2.425 & 0.980 \\
24 & 2.826 & 1.902 &	148 & 2.459 & 0.840 &	503 & 1.434 & 0.170 \\
25 & 0.002 & 0.002 &	150 & 9.085 & 0.420 &	505 & 2.008 & 1.931 \\
26 & 0.376 & 0.450 &	156 & 3.263 & 1.864 &  511 & 19.903 & 4.068 \\
28 & 4.652 & 0.990 &	163 & 1.008 & 0.340 &	516 & 0.720 & 0.668 \\
29 & 5.920 & (F) &	164 & 0.467 & 0.390 &	532 & 2.895 & 0.759 \\
31 & 3.130 & (F) &	165 & 9.761 & 0.100 &	554 & 1.575 & 1.277 \\
33 & 3.117 & 0.370 &	168 & 8.599 & 0.950 &	582 & 0.215 & 0.590 \\
34 & 1.816 & 1.703 &	173 & 6.743 & 5.668 &	584 & 0.414 & 0.290 \\
36 & 2.170 & 1.909 &	187 & 2.484 & 1.075 &	602 & 5.107 & 0.240 \\
38 & 2.872 & 2.748 &	192 & 0.719 & (F) &	604 & 0.728 & 0.140 \\
39 & 8.799 & 0.510 &	194 & 8.803 & 2.921 &	626 & 2.691 & 3.280 \\
41 & 9.213 & 2.631 &	200 & 5.387 & 0.810 &	665 & 0.351 & 0.200 \\
42 & 1.854 & 1.070 &	204 & 0.302 & 0.910 &	675 & 6.058 & 1.200 \\
43 & 0.753 & 0.430 &	210 & 1.713 & 0.550 &	679 & 0.359 & 0.100 \\
45 & 2.860 & (F) &	211 & 3.801 & 0.660 &	680 & 1.352 & 0.020 \\
46 & 3.525 & 2.630 &	212 & 6.639 & 0.530 &	690 & 6.428 & 0.140 \\
47 & 5.387 & 0.230 &	216 & 0.564 & 0.460 &	702 & 6.859 & 5.606 \\
48 & 20.106 & 14.987 &	217 & 0.765 & 0.030 &	 704 & 18.600 & (F) \\
50 & 1.833 & 1.344 &	221 & 2.951 & 0.170 &	735 & 1.081 & 0.340 \\
51 & 0.726 & 0.421 &	234 & 0.221 & 0.730 &	739 & 0.584 & 0.540 \\
52 & 42.304 & 8.056 &	240 & 2.067 & 1.912 &	747 & 6.032 & 2.306 \\
53 & 2.830 & 2.515 &	253 & 0.904 & 0.650 &	751 & 3.552 & 0.260 \\
54 & 2.237 & 1.113 &	266 & 2.085 & 0.210 &	758 & 0.468 & 0.040 \\
56 & 2.321 & 1.812 &	268 & 0.786 & 0.680 &	760 & 0.668 & 0.662 \\
59 & 2.587 & 1.795 &	304 & 0.576 & 0.563 &	769 & 3.174 & 0.320 \\
60 & 0.402 & 0.375 &	306 & 0.268 & 0.290 &	784 & 1.882 & 0.160 \\
61 & 1.454 & 1.400 &	322 & 0.936 & 0.020 &	786 & 1.416 & 1.401 \\
63 & 2.022 & 1.685 &	328 & 4.889 & 0.130 &	804 & 2.513 & 1.820 \\
65 & 7.173 & 4.256 &	337 & 0.543 & 0.080 &	914 & 1.183 & 0.120 \\
67 & 0.516 & 0.050 &	344 & 0.342 & 0.188 &	949 & 0.872 & 0.310 \\
72 & 1.668 & 4.270 &	345 & 2.195 & 0.230 &	1013 & 0.086 & 0.720 \\
74 & 3.082 & 2.696 &	346 & 3.182 & 0.090 &	1015 & 2.398 & 0.340 \\
77 & 0.874 & 0.340 &	354 & 2.451 & (F) &	1021 & 2.585 & 0.060 \\
78 & 3.232 & 2.319 &	365 & 2.937 & 0.480 &	1036 & 0.084 & 0.160 \\
81 & 3.111 & 2.671 &	372 & 4.443 & (F) &	1171 & 0.908 & 0.100 \\
\hline
\end{tabular}
\end{center}
\label{mass}
\end{table}

\begin{table}
\caption{Values of dynamical parameters fitted to LLR observations. $GM_{EMB}$ is the sum of Earth's and Moon's masses, multiplied by the gravitationnal constant and is expressed in AU\textsuperscript{3}/day\textsuperscript{2}. $C_{nmE}$ are the Earth's coefficients of potential (without unit). $\tau_{21E}$ and $\tau_{22E}$ are time delays of the Earth used for tides effects and expressed in days. $C_{nmM}$ and $S_{nmM}$ are the Moon's coefficients of potential (without unit). $(C/MR^2)_M$ is the ratio between the third moment of inertia of the Moon, divided by its mass and the square of the mean equatorial radius (without unit). $k_{2M}$ and $\tau_M$ are Love number (without unit) and time delay (in day) of the Moon. Formal errors at $1\sigma$ are given if the parameter is fitted. Fixed values come from Lunar gravity model LP15Q or Earth's one EGM96.}
\begin{center}
\begin{tabular}{c|r|l}
Name & \multicolumn{1}{c|}{Value} & \multicolumn{1}{c}{Formal error ($1\sigma$)} \\
\hline
$ GM_{EMB} $  & $ 8.9970116036\times 10^{-10} $  & $ \pm  7.2\times 10^{-19} $ \\
$ C_{20E} $  & $ -1.0826282367\times 10^{-3} $  & $ \pm  1.1\times 10^{-9} $ \\
$ C_{30E} $  & $ 2.8862851809\times 10^{-6} $  & $ \pm  3.9\times 10^{-8} $ \\
$ C_{40E} $  & $ 1.6196215914\times 10^{-6} $  &  \\
$ \tau_{21E} $  & $ 1.2326471124\times 10^{-2} $  & $ \pm  9.7\times 10^{-5} $ \\
$ \tau_{22E} $  & $ 6.9795966104\times 10^{-3} $  & $ \pm  7.7\times 10^{-6} $ \\
$ C_{20M} $  & $ -2.0347464196\times 10^{-4} $  & $ \pm  2.8\times 10^{-8} $ \\
$ C_{22M} $  & $ 2.2400822245\times 10^{-5} $  & $ \pm  2.7\times 10^{-9} $ \\
$ C_{30M} $  & $ -8.3743483591\times 10^{-6} $  & $ \pm  2.3\times 10^{-8} $ \\
$ C_{31M} $  & $ 3.1934585938\times 10^{-5} $  & $ \pm  3.7\times 10^{-7} $ \\
$ C_{32M} $  & $ 4.8452131770\times 10^{-6} $  &  \\
$ C_{33M} $  & $ 1.7262951856\times 10^{-6} $  & $ \pm  6.3\times 10^{-9} $ \\
$ C_{40M} $  & $ 9.6422863508\times 10^{-6} $  &  \\
$ C_{41M} $  & $ -5.6926874003\times 10^{-6} $  &  \\
$ C_{42M} $  & $ -1.5861997683\times 10^{-6} $  &  \\
$ C_{43M} $  & $ -8.1204110561\times 10^{-8} $  &  \\
$ C_{44M} $  & $ -1.2739414703\times 10^{-7} $  &  \\
$ S_{31M} $  & $ 3.2781092481\times 10^{-6} $  & $ \pm  8.8\times 10^{-8} $ \\
$ S_{32M} $  & $ 1.6873163963\times 10^{-6} $  & $ \pm  5.7\times 10^{-10} $ \\
$ S_{33M} $  & $ -2.4855254932\times 10^{-7} $  &  \\
$ S_{41M} $  & $ 1.5743934837\times 10^{-6} $  &  \\
$ S_{42M} $  & $ -1.5173124037\times 10^{-6} $  &  \\
$ S_{43M} $  & $ -8.0279066453\times 10^{-7} $  &  \\
$ S_{44M} $  & $ 8.3147478750\times 10^{-8} $  &  \\
$ (C/MR^2)_M $  & $ 3.9319024676\times 10^{-1} $  & $ \pm  4.7\times 10^{-5} $ \\
$ k_{2M} $  & $ 2.6248414243\times 10^{-2} $  & $ \pm  1.7\times 10^{-4} $ \\
$ \tau_{M} $  & $ 1.8896343615\times 10^{-1} $  & $ \pm  1.2\times 10^{-3} $ \\

\hline
\end{tabular}
\end{center}
\label{Tab_valeurs_parametres_LLR_dyn_I10a}
\end{table}

\begin{table}
\caption{Selenocentric coordinates of reflectors, expressed in meters.}
\begin{center}
\begin{tabular}{cc|r|l}
Reflector & &\multicolumn{1}{c|}{Value} & \multicolumn{1}{c}{Formal error ($1\sigma$)} \\
\hline
          & x  & $  1591926.238 $  &  $ \pm  1.510 $ \\
Apollo 11 & y  & $   690799.993 $  &  $ \pm  3.460 $ \\
          & z  & $    21002.457 $  &  $ \pm  0.306 $ \\
\hline
          & x  & $  1652725.404 $  &  $ \pm  1.140 $ \\
Apollo 14 & y  & $  -520893.461 $  &  $ \pm  3.600 $ \\
          & z  & $  -109731.862 $  &  $ \pm  0.318 $ \\
\hline
          & x  & $  1554675.870 $  &  $ \pm  0.276 $ \\
Apollo 15 & y  & $    98193.648 $  &  $ \pm  3.380 $ \\
          & z  & $   765004.763 $  &  $ \pm  0.304 $ \\
\hline
          & x  & $  1339316.595 $  &  $ \pm  1.750 $ \\
Lunakhod2 & y  & $   801956.645 $  &  $ \pm  2.910 $ \\
          & z  & $   756358.336 $  &  $ \pm  0.267 $ \\
\hline

\end{tabular}
\end{center}
\label{Tab_valeurs_parametres_LLR_I10a_reflecteurs}
\end{table}

\begin{table}
\caption{ITRF coordinates of stations at J1997.0, expressed in meters.}
\begin{center}
\begin{tabular}{cc|r|l}
Station & &\multicolumn{1}{c|}{Value} & \multicolumn{1}{c}{Formal error ($1\sigma$)} \\
\hline
              & x  & $  4581692.121 $  &  $ \pm  0.003 $ \\
Cerga  & y  & $   556196.025 $  &  $ \pm  0.001 $ \\
              & z  & $  4389355.020 $  &  $ \pm  0.010 $ \\
\hline
              & x  & $ -1330781.440 $  &  $ \pm  0.012 $ \\
Mc Donald  & y  & $ -5328755.514 $  &  $ \pm  0.009 $ \\
              & z  & $  3235697.496 $  &  $ \pm  0.022 $ \\
\hline
              & x  & $ -1330121.099 $  &  $ \pm  0.014 $ \\
MLRS1 & y  & $ -5328532.296 $  &  $ \pm  0.008 $ \\
              & z  & $  3236146.576 $  &  $ \pm  0.024 $ \\
\hline
              & x  & $ -1330021.431 $  &  $ \pm  0.002 $ \\
MLRS2 & y  & $ -5328403.289 $  &  $ \pm  0.003 $ \\
              & z  & $  3236481.603 $  &  $ \pm  0.010 $ \\
\hline
              & x  & $ -5466000.459 $  &  $ \pm  0.011 $ \\
Haleakala (rec.) & y  & $ -2404424.720 $  &  $ \pm  0.013 $ \\
              & z  & $  2242206.708 $  &  $ \pm  0.028 $ \\
\hline
              & x  & $ -1463998.834 $  &  $ \pm  0.005 $ \\
Apollo & y  & $ -5166632.675 $  &  $ \pm  0.004 $ \\
              & z  & $  3435013.101 $  &  $ \pm  0.012 $ \\
\hline

\end{tabular}
\end{center}
\label{Tab_valeurs_parametres_LLR_I10a_stations}
\end{table}

\begin{table}
\caption{Means and standard deviations of LLR residuals for INPOP10a solution; both are expressed in centimeters. Na is the total number of observations available, Nk is the number kept in fitting process, Nr is the number that have been rejected according to the $3\sigma$ criterion (Na=Nk+Nr).}
\begin{center}
\begin{tabular}{c|c|r|r|r|r|r}
Station & Period & Mean & Std. dev. & Na & Nk & Nr \\
\hline
Cerga      &  1987-1995  & -0.16 &   6.37 & 3460 & 3415 &   45 \\
Cerga      &  1995-2010  & -0.03 &   4.00 & 4932 & 4861 &   71 \\
Cerga      &  1984-1986  &  8.11 &  16.00 & 1187 & 1158 &   29 \\
Mc Donald  &  1969-1986  &  0.16 &  31.78 & 3604 & 3489 &  115 \\
MLRS1      &  1982-1985  & -7.89 &  73.28 &  418 &  405 &   13 \\
MLRS1      &  1985-1988  &  0.22 &   7.30 &  174 &  163 &   11 \\
MLRS2      &  1988-1996  & -0.85 &   4.27 & 1192 & 1148 &   44 \\
MLRS2      &  1996-2008  &  0.56 &   4.82 & 2243 & 1768 &  475 \\
Haleakala  &  1984-1990  & -0.44 &   8.10 &  770 &  734 &   36 \\
Apollo     &  2006-2009  &  0.08 &   4.89 &  642 &  640 &    2 \\

\hline
\end{tabular}
\end{center}
\label{Tab_residus_LLR_I10a}
\end{table}


\begin{figure}
\begin{center}
\includegraphics[scale=1]{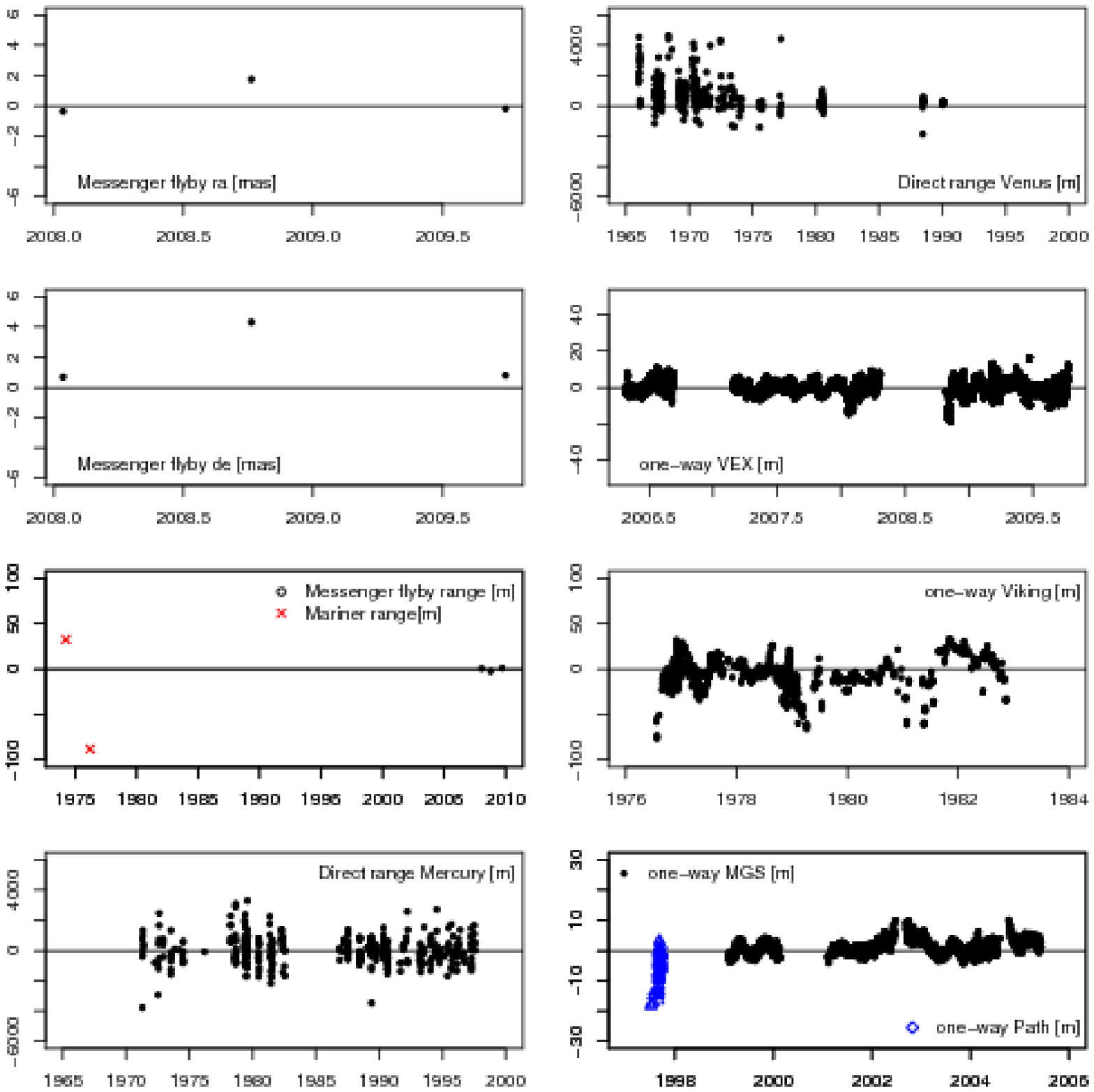} 
\caption{Postfit residuals obtained with INPOP10a}
\label{obs1}
\end{center}
\end{figure}

\begin{figure}
\begin{center}
\includegraphics[scale=1.0]{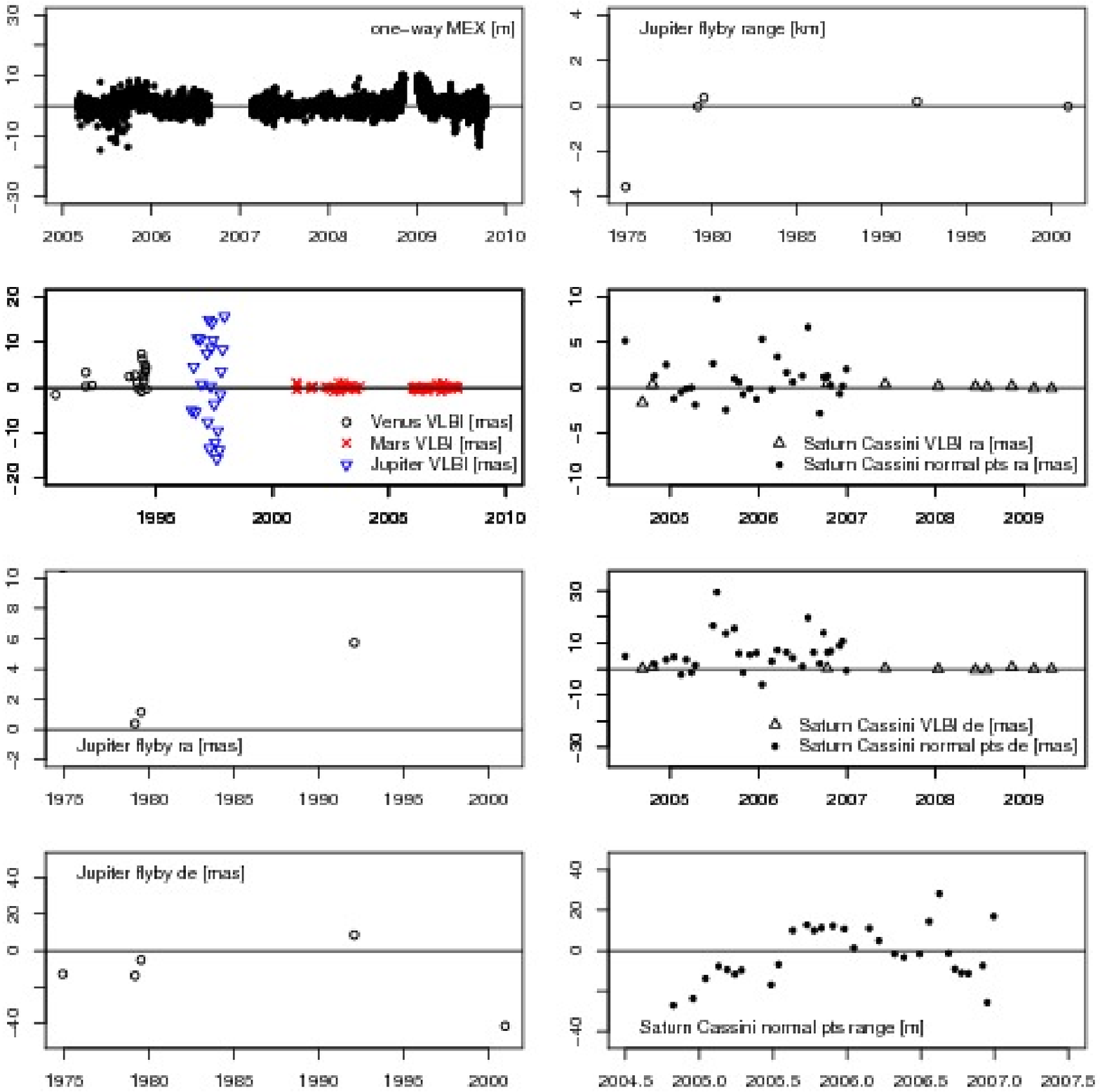} 
\caption{Postfit residuals obtained with INPOP10a}
\label{obs2}
\end{center}
\end{figure}

\begin{figure}
\begin{center}
\includegraphics[scale=1.0]{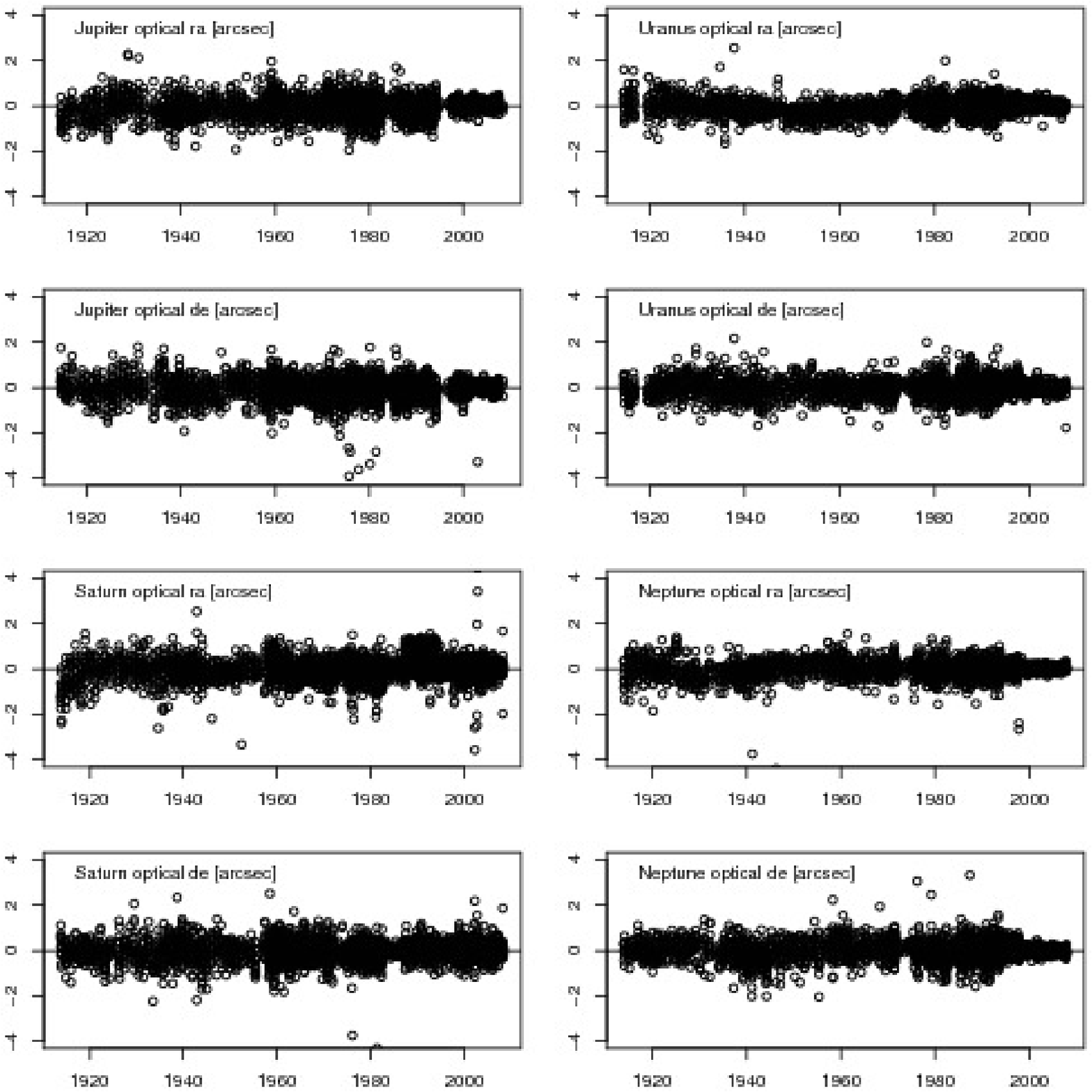} 
\caption{Postfit residuals obtained with INPOP10a}
\label{obs3}
\end{center}
\end{figure}

\begin{figure}
\begin{center}
\includegraphics[scale=0.65]{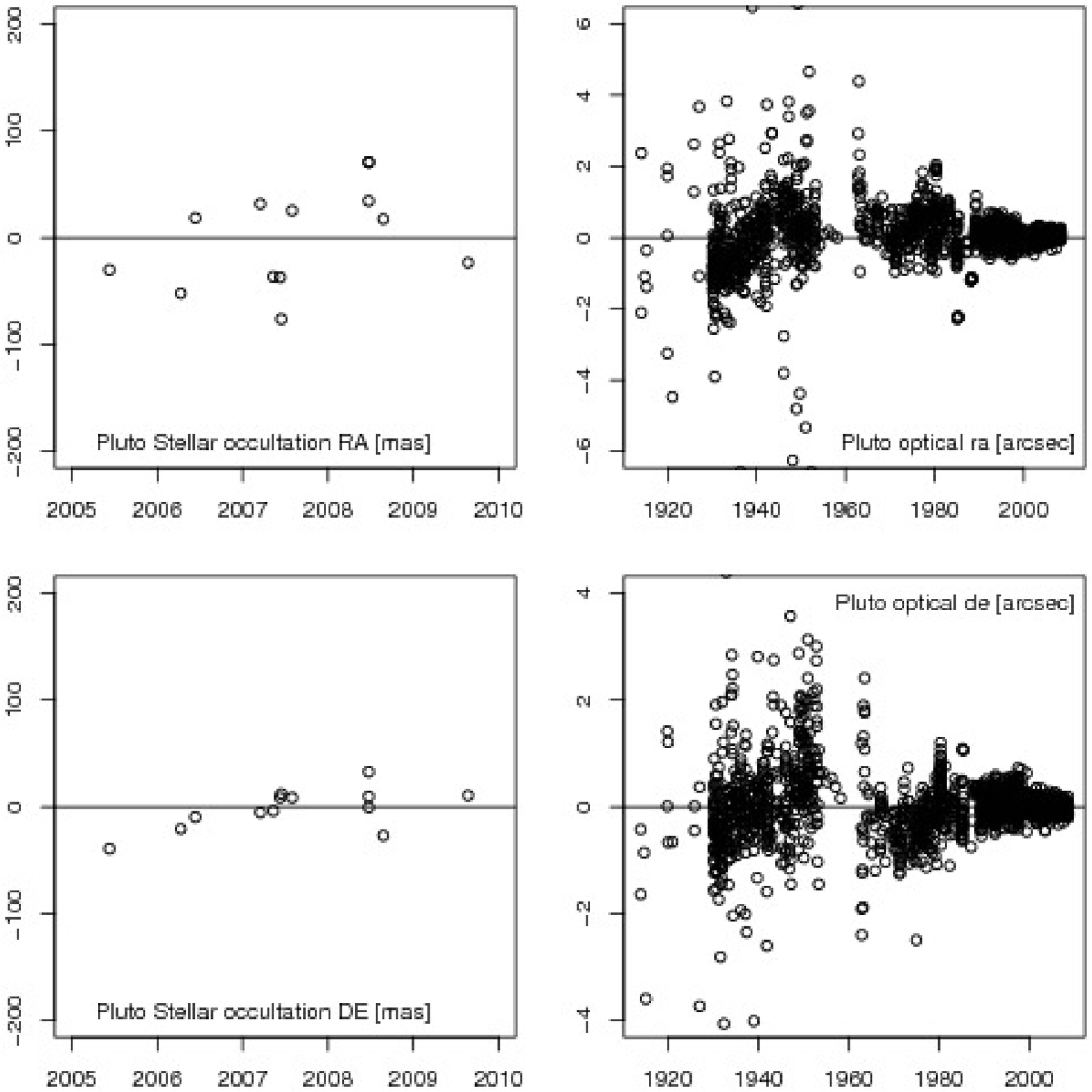} 
\caption{Postfit residuals obtained with INPOP10a}
\label{obs4}
\end{center}
\end{figure}

\clearpage


\begin{figure}
\begin{center}
\includegraphics[scale=0.45]{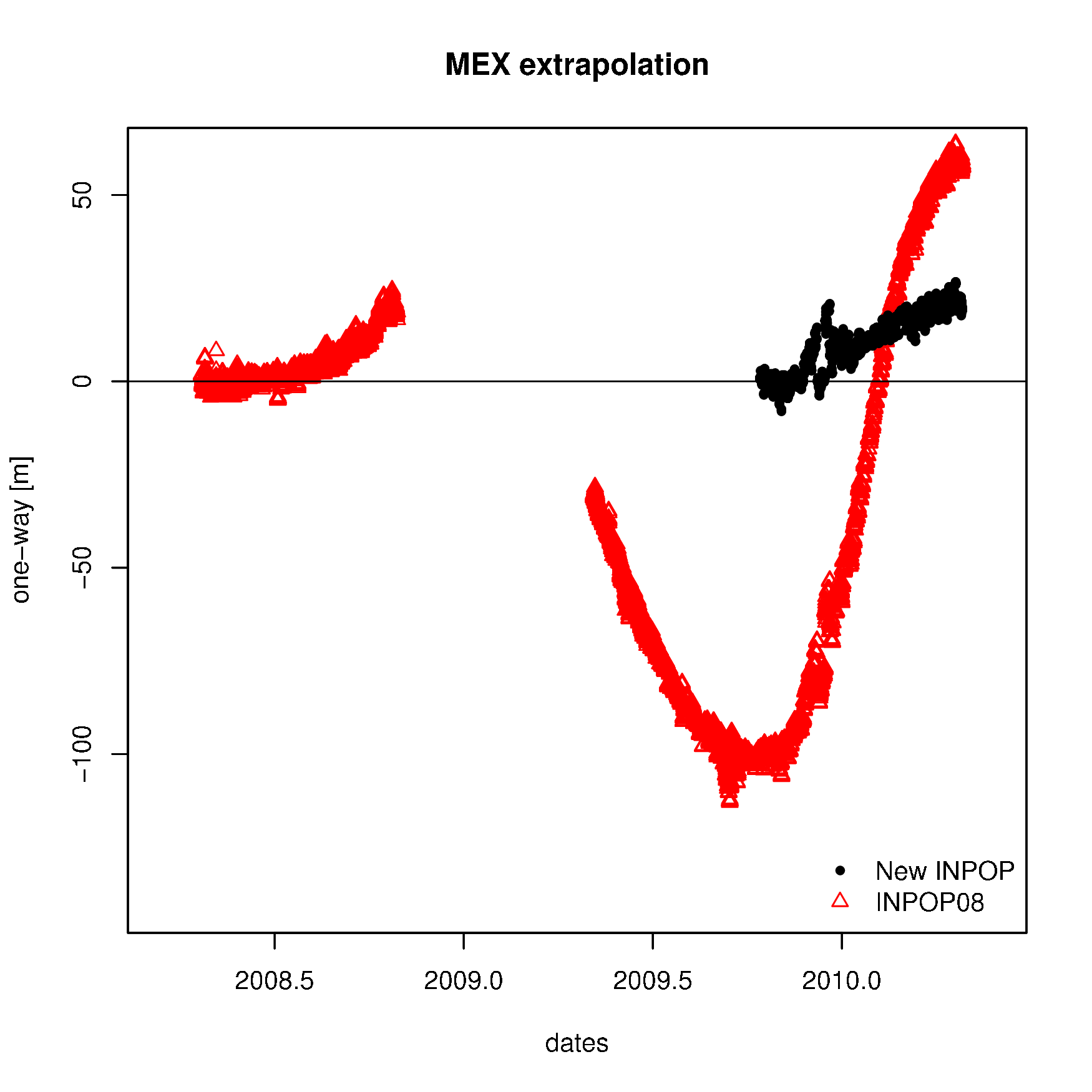}\includegraphics[scale=0.45]{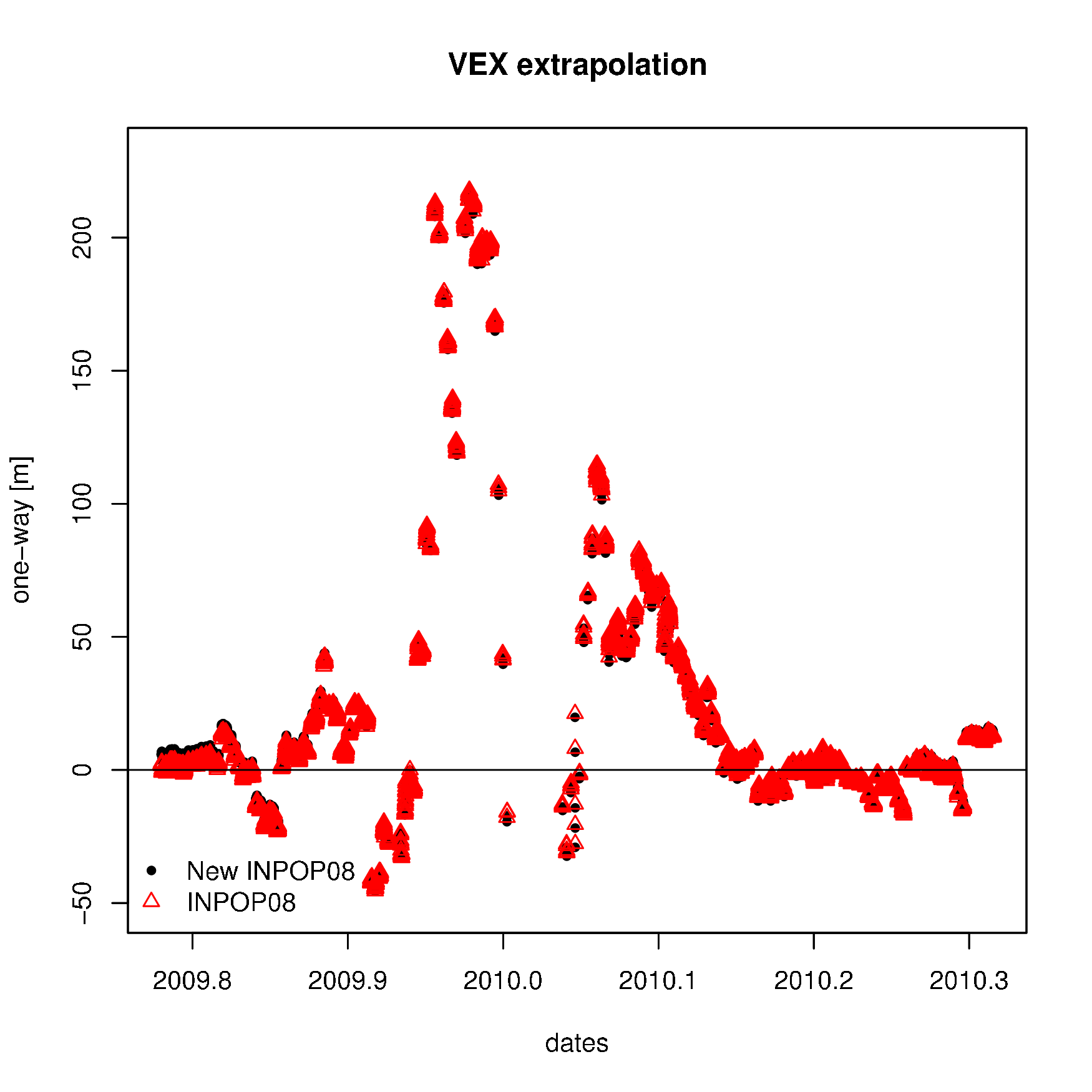} 
\end{center}
\caption{Extrapolation of MEX and VEX out from the fit interval.}
\begin{center}
\includegraphics[scale=0.45]{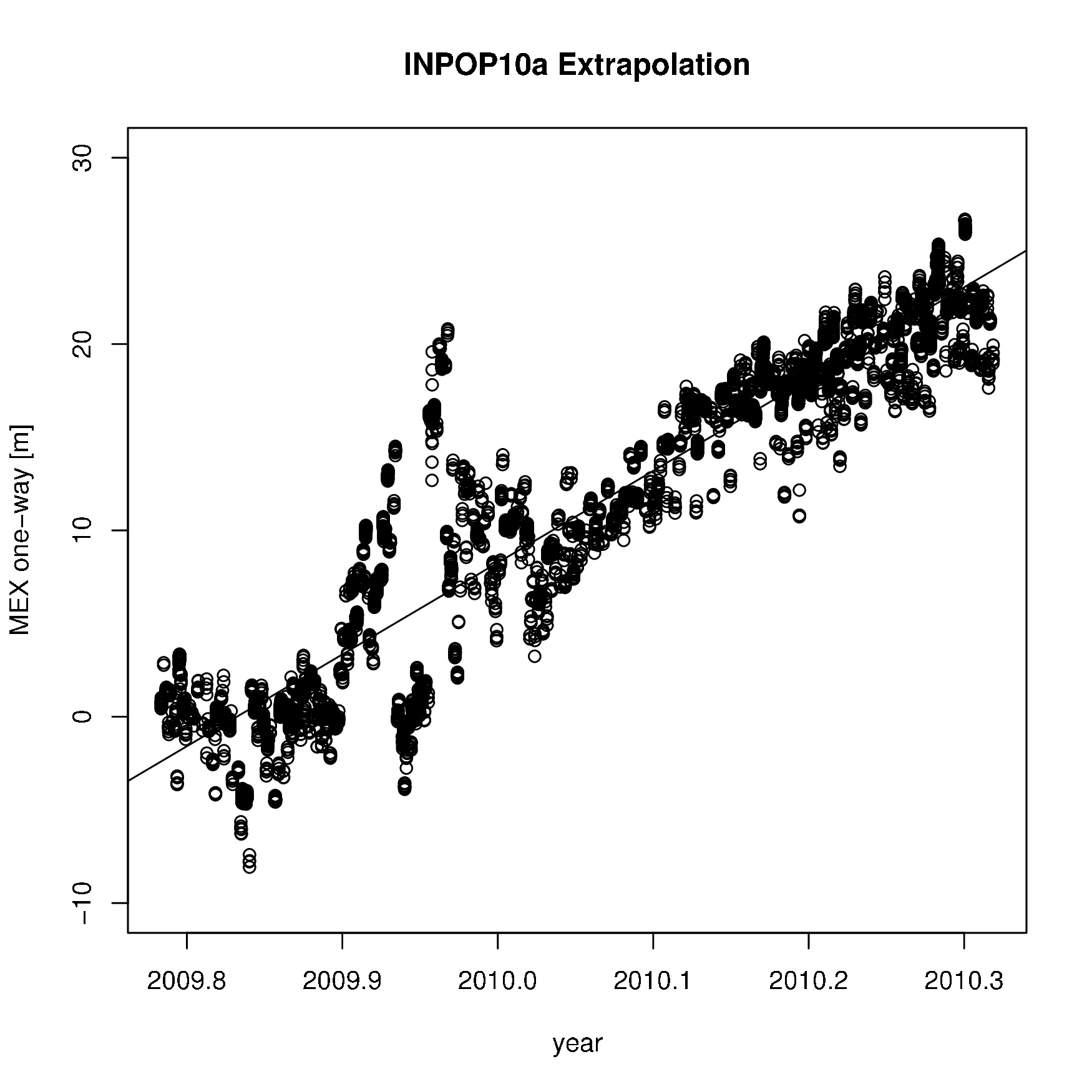}\includegraphics[scale=0.45]{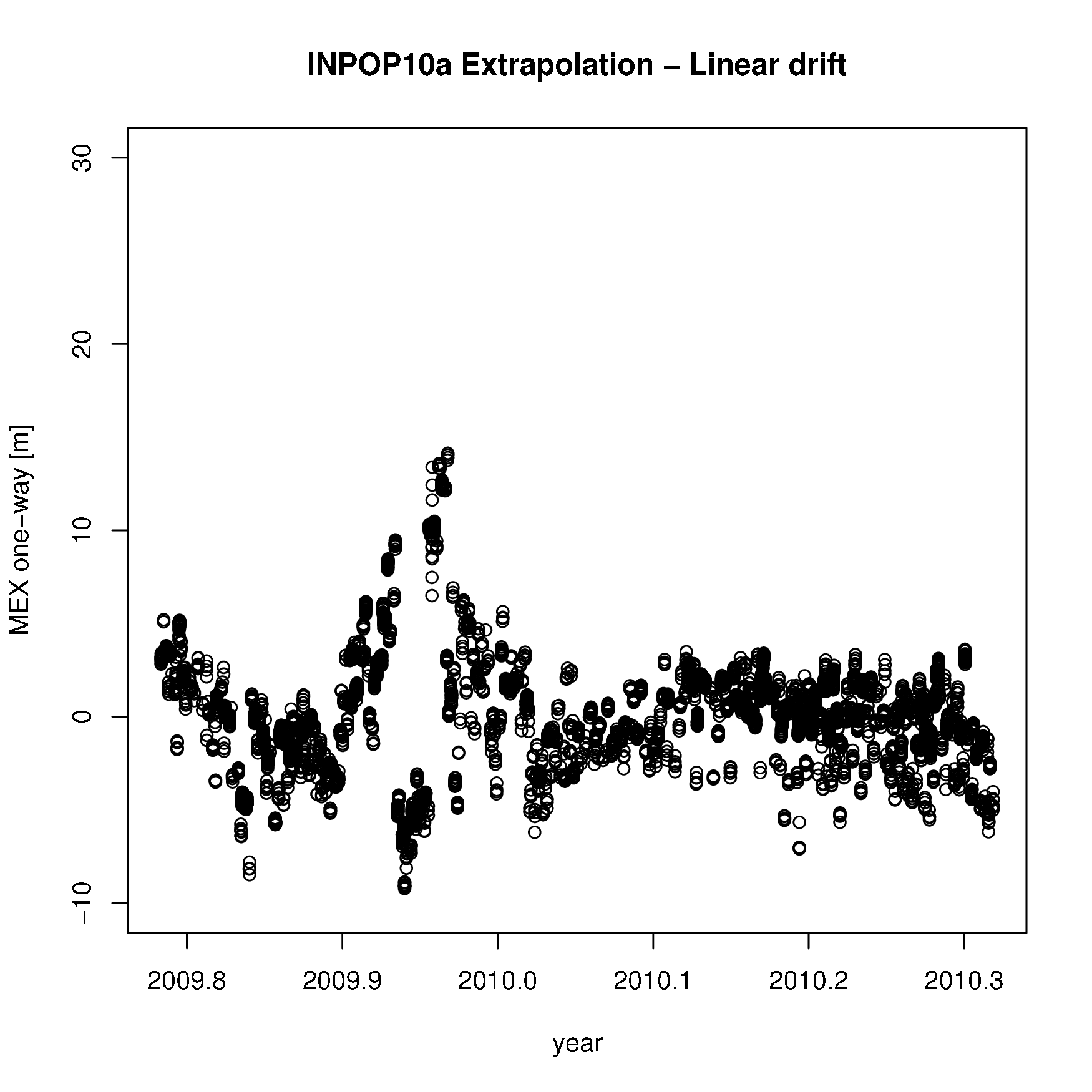}
\includegraphics[scale=0.45]{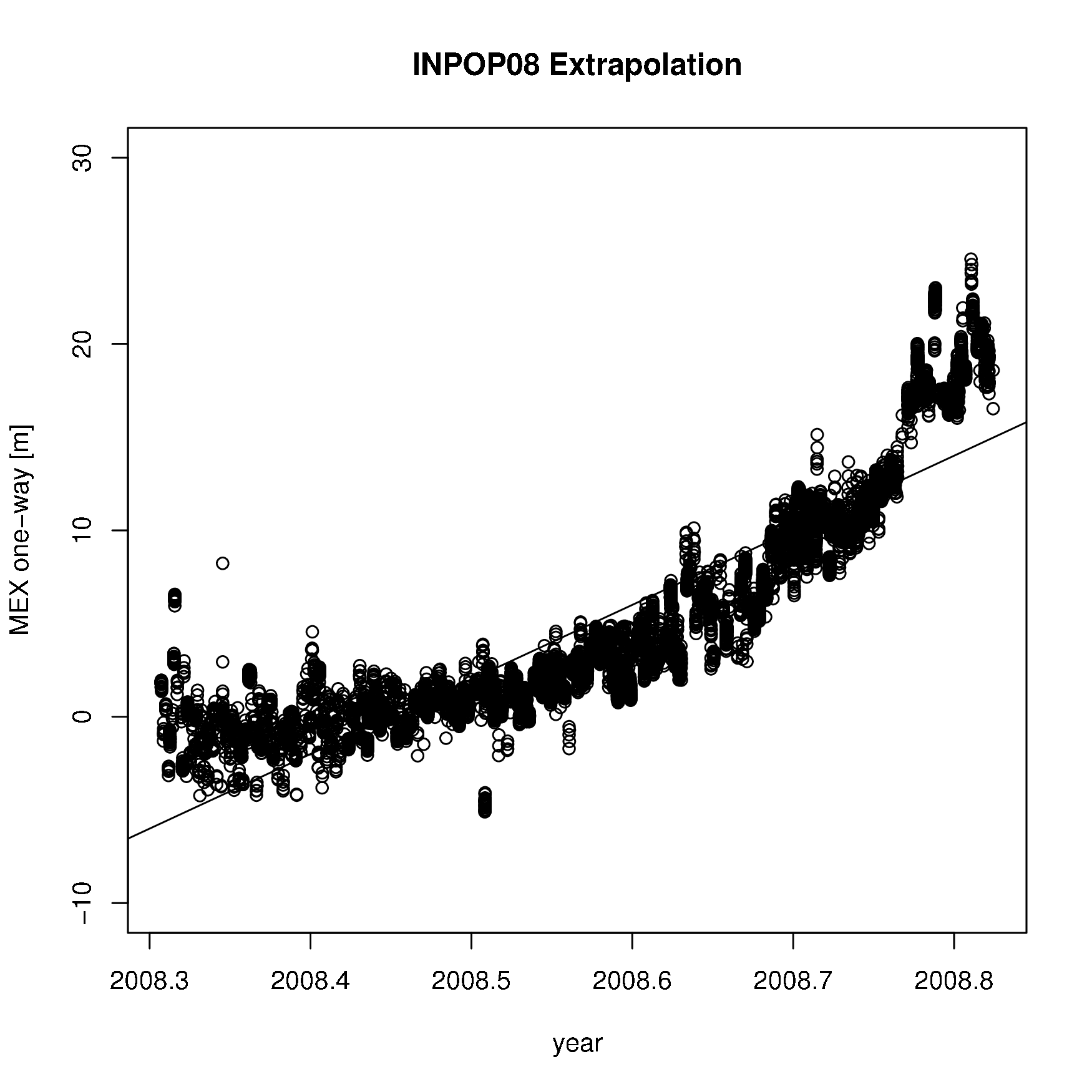}\includegraphics[scale=0.45]{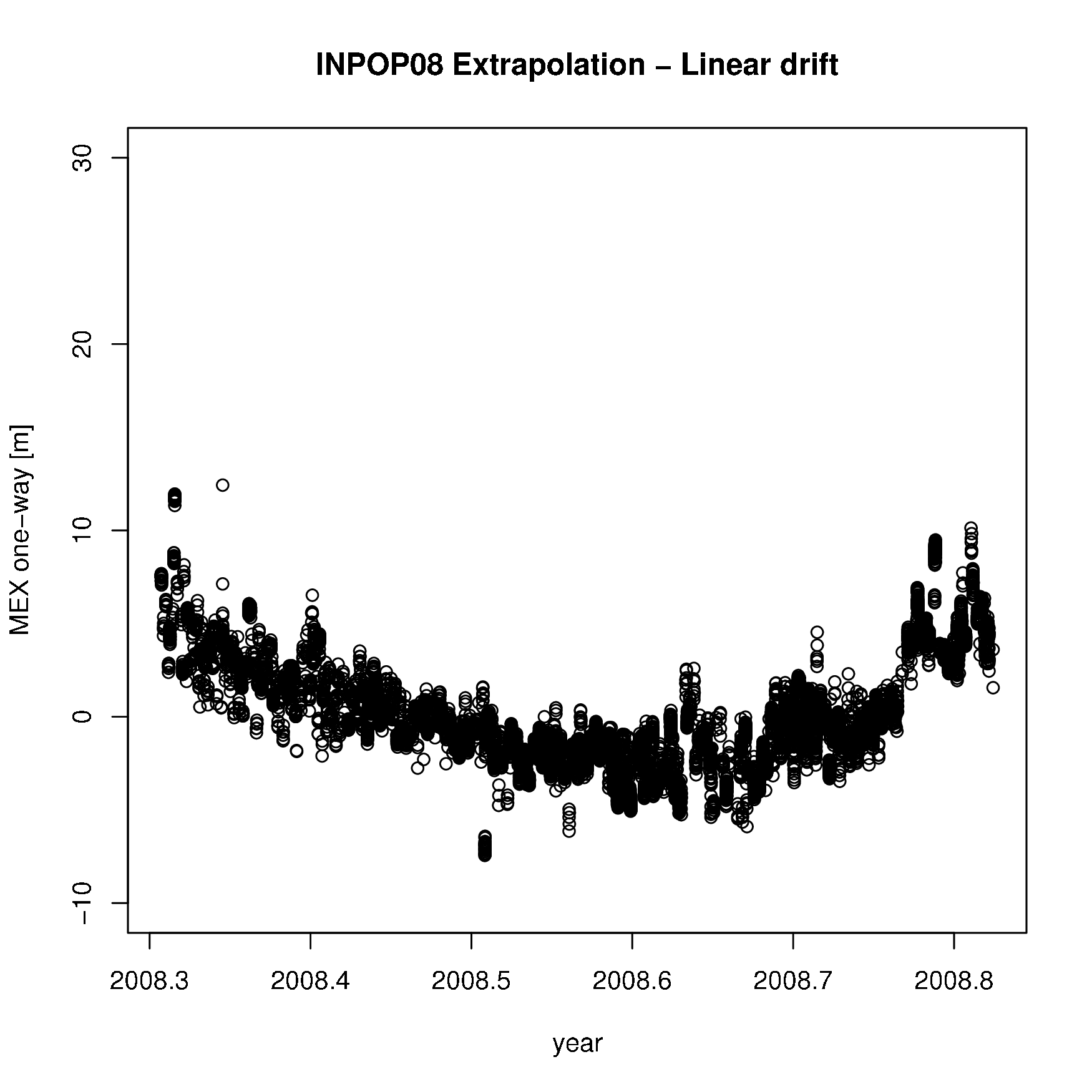}
\caption{Zooms from Figure 8 plotting the MEX extrapolated residuals.}
\end{center}
\label{omcextrapplotvol2}

\end{figure}

\clearpage

\begin{figure}
\begin{center}
\includegraphics[scale=1.0]{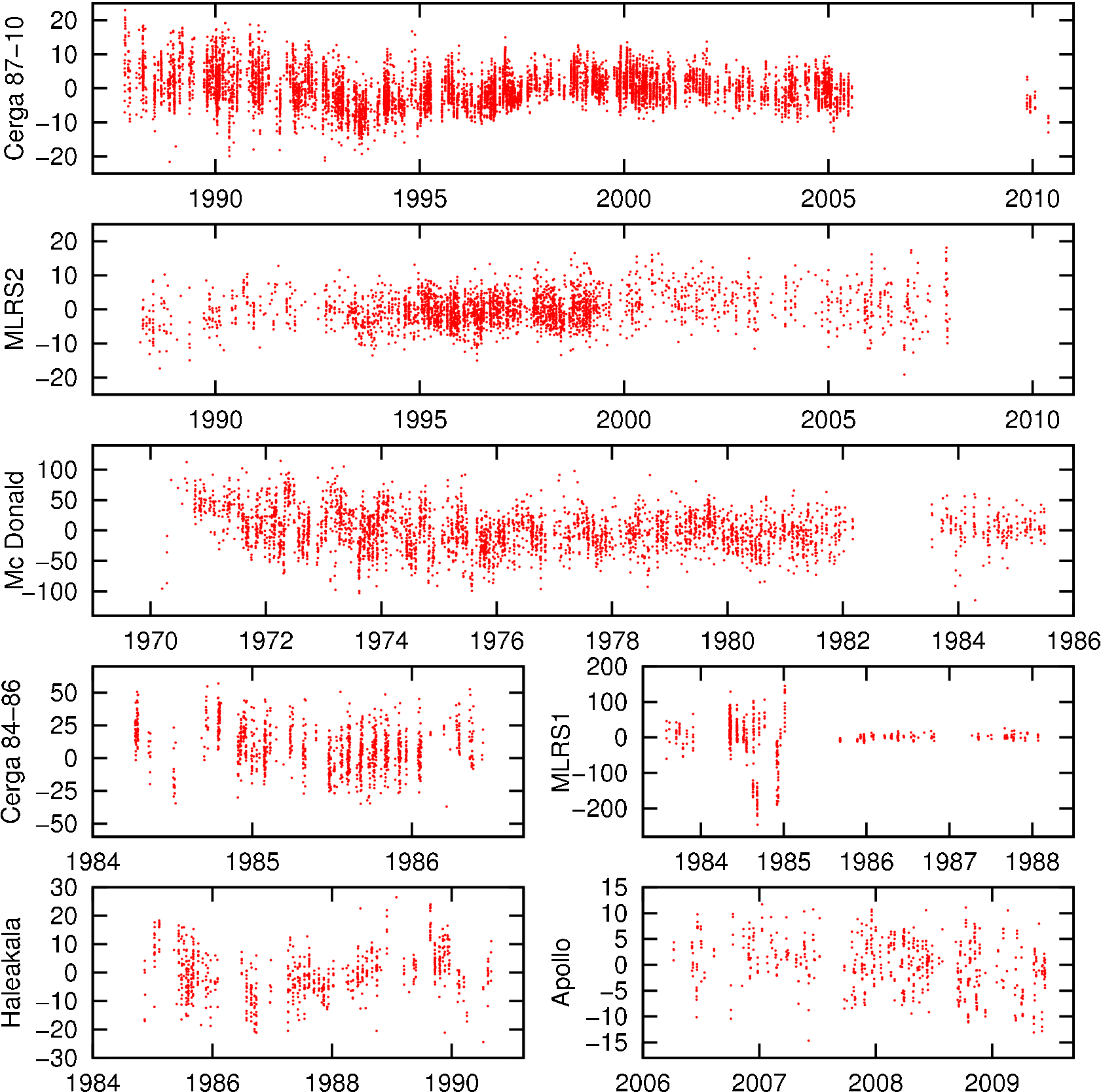} 
\caption{Postfit LLR residuals with INPOP10a for each station, expressed in centimeters.}
\label{Fig_residus_LLR_I10a}
\end{center}
\end{figure}

\end{document}